\documentstyle[12pt,axodraw]{article}
\begin{document}

\renewcommand{\theequation}{\thesection.\arabic{equation}}
\newcounter{saveeqn}
\newcommand{\add}{\addtocounter{equation}{1}}
\newcommand{\alpheqn}{\setcounter{saveeqn}{\value{equation}}%
\setcounter{equation}{0}%
\renewcommand{\theequation}{\mbox{\thesection.\arabic{saveeqn}{\alph{equation}}}}}
\newcommand{\reseteqn}{\setcounter{equation}{\value{saveeqn}}%
\renewcommand{\theequation}{\thesection.\arabic{equation}}}
\newenvironment{nedalph}{\add\alpheqn\begin{eqnarray}}{\end{eqnarray}\reseteqn}
\newsavebox{\PSLASH}
\sbox{\PSLASH}{$p$\hspace{-1.8mm}/}
\newcommand{\PS}{\usebox{\PSLASH}}
\newsavebox{\notrightarrow}
\sbox{\notrightarrow}{$\to$\hspace{-4mm}/}
\newcommand{\notto}{\usebox{\notrightarrow}}
\newsavebox{\PARTIALSLASH}
\sbox{\PARTIALSLASH}{$\partial$\hspace{-2.3mm}/}
\newcommand{\PARTIALS}{\usebox{\PARTIALSLASH}}
\newsavebox{\ASLASH}
\sbox{\ASLASH}{$A$\hspace{-2.1mm}/}
\newcommand{\AS}{\usebox{\ASLASH}}
\newsavebox{\KSLASH}
\sbox{\KSLASH}{$k$\hspace{-1.8mm}/}
\newcommand{\KS}{\usebox{\KSLASH}}
\newsavebox{\LSLASH}
\sbox{\LSLASH}{$\ell$\hspace{-1.8mm}/}
\newcommand{\LS}{\usebox{\LSLASH}}
\newsavebox{\QSLASH}
\sbox{\QSLASH}{$q$\hspace{-1.8mm}/}
\newcommand{\QS}{\usebox{\QSLASH}}
\newsavebox{\DSLASH}
\sbox{\DSLASH}{$D$\hspace{-2.8mm}/}
\newcommand{\DS}{\usebox{\DSLASH}}
\newsavebox{\DbfSLASH}
\sbox{\DbfSLASH}{${\mathbf D}$\hspace{-2.8mm}/}
\newcommand{\DBFS}{\usebox{\DbfSLASH}}
\newsavebox{\DELVECRIGHT}
\sbox{\DELVECRIGHT}{$\stackrel{\rightarrow}{\partial}$}
\newcommand{\PARVECR}{\usebox{\DELVECRIGHT}}
\thispagestyle{empty}
\begin{flushright}
IPM/P-2005/039
\end{flushright}
\vspace{0.5cm}
\begin{center}
{\large\bf{On the Anomalies and Schwinger Terms in Noncommutative Gauge Theories}}\\

\vspace{1cm} {\bf F. Ardalan, H. Arfaei} {\it and} {\bf
N. Sadooghi}  \\
\vspace{0.5cm}
{\sl Department of Physics, Sharif University of Technology}\\
{\sl P.O. Box 11365-9161, Tehran-Iran}\\
and\\
{\sl Institute for Studies in Theoretical Physics and Mathematics (IPM)}\\
{\sl{School of Physics, P.O. Box 19395-5531, Tehran-Iran}}\\
{\it E-mails: ardalan, arfaei, sadooghi@ipm.ir}
\end{center}
\vspace{0cm}
\begin{center}
{\bf {Abstract}}
\end{center}
\begin{quote}
Invariant (nonplanar) anomaly of noncommutative QED is reexamined. It is found that just as in ordinary gauge
theory UV regularization is needed to discover anomalies, in noncommutative case, in addition, an IR
regularization is also required to exhibit existence of invariant anomaly. Thus resolving the controversy in the
value of invariant anomaly, an expression for the unintegrated anomaly is found. Schwinger terms of the current
algebra of the theory are derived.

\end{quote}
\newpage
\newpage
\subsubsection*{1\hspace{0.3cm}Introduction}
\setcounter{section}{1} \setcounter{equation}{0} One of the
greatest achievements of theoretical physics in the second half of
the last century is the discovery of the breakdown of classical
symmetries of field theories due to quantum effects
\cite{adler-old1, adler-old2, anomaly-book, adam-new}. The far
reaching consequences of this discovery include quantitative
predictions of physical amplitudes from anomaly in global
symmetries such as in two photon decay of pions, and restriction
of consistent gauge theory models of particle physics from
cancellation of anomalies in local symmetries such as in
electroweak theory (for recent review of anomaly see
\cite{adler-new} and references therein).
\par
Therefore, in study of various field theories, it is essential to
understand anomalies of its various classical global and local
symmetries. One class of field theories studied extensively
recently is noncommutative field theories, in particular gauge
theories. Anomalies of noncommutative gauge theories have been
widely studied \cite{neda1, neda2, martin2, intriligator, armoni,
ncanomalies2} and are mostly well understood.
\par
In ordinary commutative gauge theories, currents corresponding to {\it global symmetries} satisfy an ordinary
divergence equation
\begin{eqnarray*}
\partial_{\mu}j^{\mu}_{5}=0,
\end{eqnarray*}
while chiral currents corresponding to {\it local symmetries} satisfy a covariant divergence equation
\begin{eqnarray*}
D_{\mu}J^{\mu}=0,
\end{eqnarray*}
where $D_{\mu}=\partial_{\mu}-igA_{\mu}$, with $A_{\mu}$ acting in
the adjoint representation of the gauge group. Generally these two
divergence equations receive nonzero contributions from quantum
correction.
\par
In the noncommutative gauge theories a similar situation exists.
For simplicity we restrict our attention to a $U(1)$
noncommutative gauge theory, as noncommutativity already
incorporates complications inherent in the non-Abelian nature of
gauge theories with larger gauge groups.
\par
In this case, there are two distinct global Noether currents which
play the role of the two currents above, the invariant
$j_{\mu}^{5}\equiv
\bar{\psi}_{\alpha}\star\psi_{\beta}(\gamma_{\mu}\gamma_{5})^{\alpha\beta},$
and the covariant current
$J_{\mu}^{5}\equiv
\psi_{\beta}\star\bar{\psi}_{\alpha}(\gamma_{\mu}\gamma_{5})^{\alpha\beta}.
$
These two currents correspond to the same global $U(1)$ symmetry
and have divergence equations
\begin{eqnarray}\label{A1}
\partial_{\mu} j^{\mu}_{5}=0,
\end{eqnarray}
and
\begin{eqnarray}\label{A2}
D_{\mu} J^{\mu}_{5}=0,
\end{eqnarray}
where $D_{\mu}$ acts in the $\star$-adjoint operation to be
defined in section 2.
\par
The anomaly of the covariant conservation equation (\ref{A2}) has
been unequivocally calculated using various UV regularization
methods in \cite{neda1, martin2, ncanomalies2, ncanomalies3,
neda3} and is
\begin{eqnarray}\label{A3}
D_{\mu}J^{\mu}_{5}(x)=-\frac{g^{2}}{16\pi^{2}}F_{\mu\nu}(x)\star
\tilde{F}^{\mu\nu}(x).
\end{eqnarray}
The anomaly contribution to the invariant current $j^{\mu}_{5}$ is
however less uncontroversial  \cite{neda2, intriligator, armoni,
neda3}.
\par
In fact anomaly of $\partial_{\mu}j^{\mu}_{5}$ was calculated with
a UV cutoff in the spirit of UV/IR mixing phenomenon
\cite{minwalla} in \cite{neda2} and found to be
\begin{eqnarray}\label{A4}
\partial_{\mu}j^{\mu}_{5}(x)=-\frac{g^{2}}{16\pi^{2}}F_{\mu\nu}(x)\star'
\tilde{F}^{\mu\nu}(x)+\cdots,
\end{eqnarray}
for the limit of small noncommutative momentum relative to the UV
cutoff; while it was observed in \cite{martin2} that the
contribution to this conservation equation involve only nonplanar
diagrams which are convergent and therefore need no UV
regularization, leading to zero anomaly.
\par
This result was confirmed in \cite{intriligator} on the basis of
string theoretical consideration and suggested to be the
consequence of Green-Schwarz mechanism operating autonomously in
field theory. Later authors of \cite{armoni} pointed out that as
the two currents $J_{5}^{\mu}$ and $j_{5}^{\mu}$ lead to the same
global symmetry and charge
\begin{eqnarray}
Q_{5}=\int d^{3}x J^{0}_{5}(x)=\int d^{3}x j^{0}_{5}(x),
\end{eqnarray}
 and as
$J_{5}^{\mu}$ has an anomaly, then $j_{5}^{\mu}$ must be also
anomalous and in fact
\begin{eqnarray}\label{A1-3}
 \int d^{2}x_{NC}\
D_{\mu}J^{\mu}_{5}(x)=\int d^{2}x_{NC}\
\partial_{\mu}j^{\mu}_{5}(x).
\end{eqnarray}
These authors proposed to resolve this paradox by quoting Ref.
\cite{neda2} that in the momentum representation, after taking the
infinite limit of the UV cutoff, anomaly is zero everywhere except
at the zero momentum in the noncommutative directions. They made
an independent calculation in the point-splitting method and
confirmed that their zero momentum point anomaly is finite. Yet,
close scrutiny of the calculation reveals that the anomaly in the
coordinate representation gives zero as it is zero everywhere in
the momentum representation except for a finite value at zero
momentum, this having measure zero in the Lebesgues measure.
\par
Thus the paradox remains. The authors of Ref. \cite{armoni}
realize this difficulty where they take the infinitesimal
parameter of gauge transformation in the variation of the
effective action, to be a Dirac $\delta$-function in momentum
variables.
\par
Our proposal is that just as in the ordinary commutative theory
anomaly will not appear unless a UV regularization\footnote{It is
sometimes possible to obtain anomalies as IR phenomenon
\cite{anomaly-IR}.} is introduced and careless manipulation of
divergent integrals are avoided, so it is in the noncommutative
gauge theory: In addition to the UV regularization, an IR
regularization must also be introduced to exhibit a finite anomaly
in the invariant current divergence $\partial_{\mu}j^{\mu}_{5}$,
thus resolving the above paradox. For this regularization we use
compactification of the noncommutative directions\footnote{Similar
compactification in other contexts of noncommutative field theory
were performed in \cite{compactification}.}  with radius $R$. We
find an unintegrated form of anomaly for finite $R$
$$\partial_{\mu}j^{\mu}_{5}=-\frac{1}{(2R)^{2}}\ \frac{g^{2}}{16\pi^{2}}\int\limits_{-R}^{+R} d^{2}x_{NC}
F_{\mu\nu}\tilde{F}^{\mu\nu}.$$
Integrating both sides over noncommutative directions, the $R$
dependence on the r.h.s. cancels out, satisfying the constraint
equation (\ref{A1-3}) for $R \to \infty$ limit. The resulting
integrated form of the anomaly is in accordance with Ref.
\cite{armoni}. of Armoni, Lopez and Theisen. On the other hand,
taking the decompactification limit first and then integrating
over noncommutative directions yields a zero anomaly. In agreement
with Ref. \cite{intriligator} of Intriligator and Kumar.
\par
We will then use the same technique to calculate the current
commutators and obtain Schwinger terms reminiscent of the central
charge of the affine algebra appearing in conformal field
theories.
\par
There has been another calculation of the invariant anomaly
\cite{ncanomalies3} which gives an unintegrated result on which we
will briefly comment later.
\par
In the next section, we set down our notation and briefly review
the various calculations of axial anomalies for the commutative
and noncommutative gauge theories. In section 3, we make careful
point-splitting calculation in the presence of both a UV and an IR
regulators.
\par
In section 4, we present our results for current algebra of the
theory and section 5 is devoted to discussion where string
theoretical aspects are touched upon.
\subsubsection*{2\hspace{0.3cm} Background on Anomalies}
\setcounter{section}{2} \setcounter{equation}{0} The simplest and
earliest and the most thoroughly studied classical symmetry broken
by quantum corrections is the axial symmetry
\begin{eqnarray}\label{A2-1}
\psi(x)\to e^{i\gamma_{5}\alpha}\psi(x),
\end{eqnarray}
of massless QED
\begin{eqnarray}\label{A2-2}
{\cal{L}}=\bar{\psi}\left(i\PARTIALS-g\AS\right)\psi-\frac{1}{4}F_{\mu\nu}F^{\mu\nu},
\end{eqnarray}
with its classical current conservation
\begin{eqnarray}\label{A2-3}
\partial_{\mu}j^{\mu}_{5}=0,\qquad\qquad
j_{5}^{\mu}=\bar{\psi}\gamma^{\mu}\gamma^{5}\psi,
\end{eqnarray}
broken by the anomaly \cite{adler-old1}
\begin{eqnarray}\label{A2-4}
\langle\partial_{\mu}j^{\mu}_{5}\rangle=-\frac{g^{2}}{16\pi^{2}}F\wedge
F\equiv -\frac{g^{2}}{16\pi^{2}}\varepsilon_{\mu\nu\lambda\nu}
F^{\mu\nu}F^{\lambda\nu}.
\end{eqnarray}
This effect was first discovered through careful analysis of the
divergent triangle diagram contributions to the vacuum expectation
value of the current
\begin{eqnarray}\label{A2-7}
\langle\partial_{\mu}j^{\mu}_{5}(x)\rangle=\int d^{4}y\ d^{4}z \
\partial_{\mu}\Gamma^{\mu\lambda\nu}(x,y,z)A_{\lambda}(y)A_{\nu}(z),
\end{eqnarray}
in which  the divergence of the three-point function
$\Gamma^{\mu\lambda\nu}$ is the sum of two divergent integrals of
the cross diagrams (see Figure 1)
\begin{eqnarray}\label{A2-7a}
\lefteqn{\hspace{-2cm}\partial_{\mu}\Gamma^{\mu\lambda\nu}(x,y,z)\equiv
\partial^{\mu}_{x}\langle \mbox{T}(
j_{5}^{\mu}(x)j^{\lambda}(y)j^{\nu}(z))\rangle=-g^{2}\int
\frac{d^{4}k}{(2\pi)^{4}}\ \frac{d^{4}p}{(2\pi)^{4}}\
e^{-i(k+p)x}e^{iky}e^{ipz}
}\nonumber\\
&&\times \int\frac{d^{4}\ell}{(2\pi)^{4}}\
\mbox{tr}\left(\gamma^{\mu}\gamma^{5}\frac{(\LS-\KS)}{(\ell-k)^{2}}\gamma^{\lambda}\frac{\LS}{\ell^{2}}
\gamma^{\nu}\frac{(\ell+\PS)}{(\LS+p)^{2}}\right)+\left(\lambda\leftrightarrow\nu,k\leftrightarrow
p\right),
\end{eqnarray}
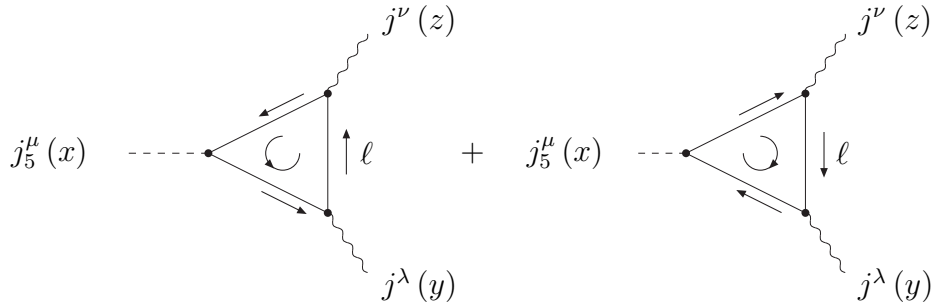
\begin{figure}[h]
\SetScale{0.3}
\begin{center}
\begin{picture} (200,40)(0,20)
\Vertex(0,0){5} \Vertex(150,75){5} \Vertex(150,-75){5}
\Line(0,0)(150,75) \Line(0,0)(150,-75) \Line(150,75)(150,-75)
\DashLine(-100,0)(0,0){10} \Photon(150,75)(200,150){3}{4}
\Photon(150,-75)(200,-150){3}{4} \Text(60,0)[]{$\ell$}
\Text(-60,0)[]{$j^{\mu}_{5}\left(x\right)$}
\Text(80,50)[]{$j^{\nu}\left(z\right)$}
\Text(80,-50)[]{$j^{\lambda}\left(y\right)$} \SetScale{0.8}
\LongArrow(65,-10)(65,10) \LongArrow(45,28)(25,18)
\LongArrow(25,-18)(45,-28) \ArrowArc(35,0)(8,90,360)
\SetScale{0.3}
\Text(100,0)[]{$+$}
\Vertex(600,0){5} \Vertex(750,75){5} \Vertex(750,-75){5}
\Line(600,0)(750,75) \Line(600,0)(750,-75) \Line(750,75)(750,-75)
\DashLine(540,0)(600,0){10} \Photon(750,75)(800,150){3}{4}
\Photon(750,-75)(800,-150){3}{4}
\Text(240,0)[]{$\ell$} \Text(135,0)[]{$j^{\mu}_{5}\left(x\right)$}
\Text(260,50)[]{$j^{\nu}\left(z\right)$}
\Text(260,-50)[]{$j^{\lambda}\left(y\right)$} \SetScale{0.8}
\LongArrow(290,10)(290,-10) \LongArrow(250,18)(270,28) \LongArrow
(270,-28)(250,-18) \ArrowArcn(260,0)(8,90,180)
\end{picture}
\end{center}
\vskip2.5cm \caption{Triangle diagrams giving rise to the axial
anomaly.}
\end{figure}
which would cancel to give zero anomaly if a simple shift of
integration is performed.
\par\noindent
However, as the integrals are linearly divergent such change of
variable may not be admissible. After UV regularization of the
integrals the nontrivial anomaly of (\ref{A2-4}) arises.
\par
This same result is obtained by another UV regularization of
point-splitting of the poorly defined operator product
$\bar{\psi}\psi$
\begin{eqnarray}\label{A2-8}
j_{5}^{\mu}(x)=\lim\limits_{\epsilon\to
0}\bar{\psi}(x+\frac{\epsilon}{2})\gamma^{\mu}\gamma^{5}\
\exp\left({ig\int\limits_{x-\frac{\epsilon}{2}}^{x+\frac{\epsilon}{2}}A_{\mu}dx^{\mu}}\right)\psi(x-\frac{\epsilon}{2}).
\end{eqnarray}
A more elegant but less rigorous calculation of anomaly was
obtained by the observation that in the path integral calculation
of the vacuum expectation value of the axial vector current, the
measure of the path integral of the partition function is not
invariant under the map (\ref{A2-1}); therefore the divergence of
the current $j_{5}^{\mu}$ obtains a quantum correction which may
be calculated with a UV regularization of the path integral
measure Jacobian \cite{fujikawa}.
\par\noindent
The method we use in this work, the point-splitting method, will
be dealt with in detail for a noncommutative theory in section 3.
\par
Generalization of the anomaly phenomenon to non-Abelian gauge theories has interesting nontrivial results. For
one, in calculating quantum corrections to the classical conservation equation of {\it global axial flavor
symmetries}, higher order Green functions (square and pentagon diagrams) must be taken into account giving the
same result as
\begin{eqnarray}\label{A2-9}
\partial_{\mu}j^{\mu,a}_{5}=-\frac{g^{2}}{16\pi^{2}}D^{abc} F^{b}\wedge F^{c}\equiv -\frac{g^{2}}{16\pi^{2}}D^{abc}\varepsilon^{\mu\nu\rho\lambda}
F^{b}_{\mu\nu}(x)F^{c}_{\rho\lambda}(x),
\end{eqnarray}
where $D^{abc}\equiv \mbox{tr}\left(t^{a}\{t^{b},t^{c}\}\right)$; with $t^{b}, t^{c}$ the generators of the
non-Abelian gauge theory in the representation of the fermions circulating in the loop, and $t^{a}$ the
generator of the global flavor symmetry. It is these higher order diagrams that make (\ref{A2-9}) gauge
invariant \cite{weinberg}.\footnote{Here, we have distinguished between the current
$j_{a}^{\mu}=\bar{\psi}\gamma^{\mu}t_{a}\psi$, which plays the role of a {\it global flavor symmetry current}
when $t_{a}$ corresponds to {\it global} symmetries of the theory and a {\it non-Abelian gauge current} when
$t_{a}$ corresponds to the generator of the {\it local gauge} symmetry of the theory. Whereas the former
satisfies ordinary classical conservation equation $\partial_{\mu}j^{\mu,5}_{a}=0$ and is gauge invariant, the
latter satisfies classically the covariant current conservation $D_{\mu}j^{\mu}_{5}=0$ and is gauge covariant.}
\par
For another, the {\it (chiral) local gauge symmetry current} which classically satisfies the covariant
divergence equation
\begin{eqnarray}\label{A2-11}
D_{\mu}J^{\mu}=0,
\end{eqnarray}
now receives a quantum correction of the form
\begin{eqnarray}\label{A2-12}
D^{\mu}J^{a}_{\mu}=-\frac{i}{12\pi^{2}}\varepsilon^{\mu\nu\rho\sigma}\mbox{tr}\left(t^{a}\partial^{\mu}
\left(A_{\nu}\partial_{\rho}A_{\sigma}-\frac{i}{2}A_{\nu}A_{\rho}A_{\sigma}\right)\right),
\end{eqnarray}
satisfying the Wess-Zumino consistency conditions
\cite{adler-old2, anomaly-book, adam-new}, where $t^{a}$ is also a
generator of the gauge group.
\par
These quantum violations of symmetries occur for noncommutative
gauge theories also, but with some surprises. For simplicity, we
limit ourselves to the noncommutative version of massless QED.
Noncommutative massless QED has the same Lagrangian as
(\ref{A2-2})
\begin{eqnarray}\label{A2-13}
{\cal{L}}=\bar{\psi}\star\left(i\PARTIALS-g\AS\right)\star\psi-\frac{1}{4}F_{\mu\nu}\star
F^{\mu\nu},
\end{eqnarray}
with product of function replaced by $\star$-product
\begin{eqnarray}\label{A2-14}
(f\star g)(x)\equiv f(x+\xi)\
\exp\left(\frac{i\Theta^{\mu\nu}}{2}\
\frac{\partial}{\partial\xi^{\mu}}\frac{\partial}{\partial\zeta^{\nu}}\right)
g(x+\zeta)\Bigg|_{\xi=\zeta=0},
\end{eqnarray}
{\it e.g.}
\begin{eqnarray}\label{A2-15}
e^{ikx}\star e^{ipx}=e^{i(k+p)x}e^{ik\wedge p},
\end{eqnarray}
with $k\wedge p=\Theta_{\mu\nu}k^{\mu}p^{\nu}$. From now on, we
limit ourselves to the noncommutativity between space coordinates
$x_{1}$ and $x_{2}$. Obviously $\star$-product is not a
commutative operation and therefore noncommutative QED closely
resembles an ordinary non-Abelian gauge theory. In fact in
(\ref{A2-13})
\begin{eqnarray}\label{A2-16}
F_{\mu\nu}=\partial_{\mu}A_{\nu}-\partial_{\nu}A_{\mu}+ig\big[A_{\mu},A_{\nu}\big]_{\star}.
\end{eqnarray}
Now noncommutative massless QED again has an axial $U(1)$ symmetry
(\ref{A2-1}), $\psi\to e^{i\alpha\gamma_{5}}\psi$, to which now
corresponds two currents
\begin{eqnarray}\label{A2-17}
J_{\mu}^{5}\equiv
\psi_{\beta}\star\bar{\psi}_{\alpha}(\gamma_{\mu}\gamma_{5})^{\alpha\beta},
\end{eqnarray}
and \begin{eqnarray}\label{A2-18}
j_{\mu}^{5}\equiv
\bar{\psi}_{\alpha}\star\psi_{\beta}(\gamma_{\mu}\gamma_{5})^{\alpha\beta}.
\end{eqnarray}
Under a $\star$-gauge transformation $ \psi(x)\to
e^{i\alpha(x)}\star \psi(x)$, these two currents are covariant and
invariant, respectively. They have the same charge operator
\begin{eqnarray}\label{A2-19}
Q_{5}=\int d^{3}x j_{5}^{0}(x)=\int d^{3}x J^{0}_{5}(x).
\end{eqnarray}
Naively, one would expect the two currents to have the same
anomaly. But, while
\begin{eqnarray}\label{A2-20}
D_{\mu}J^{\mu}_{5}=-\frac{g^{2}}{16\pi^{2}}F_{\mu\nu}\star\tilde{F}^{\mu\nu},
\end{eqnarray}
the anomaly for the divergence of the ``invariant'' current
$j_{5}^{\mu}$ is quite different. The reason for the difference of
behavior of the two currents is an important property of
noncommutative field theory, UV/IR mixing, which shows up in the
''nonplanar'' diagrams of the theory. Simply put, UV divergences
of loop integrals, upon regularization, turn into IR divergences
in the variable $\tilde{p}^{\mu}\equiv \Theta^{\mu\nu} p_{\nu}$.
\par
In fact the first indication of nonvanishing anomaly in invariant
current $j_{5}^{\mu}$ came from this source and emerges from the
IR divergence of the triangle diagram integral \cite{neda2}.
Quoting the result from \cite{neda2}, the divergence of the
invariant axial vector current is
\begin{eqnarray}\label{A2-21}
\lefteqn{\langle\partial_{\mu}j^{\mu}_{5}(x)\rangle=-\frac{1}{\pi^{2}}
\varepsilon_{\lambda\nu\alpha\beta}\int
\frac{d^{4}k}{(2\pi)^{4}}\frac{d^{4}p}{(2\pi)^{4}}\
k^{\alpha}A^{\lambda}(k)e^{-ikx}p^{\beta}A^{\nu}(p)e^{-ipx}}\nonumber\\
&&\times\int\limits_{0}^{1}d\alpha_{1}\int\limits_{0}^{1-\alpha_{1}}d\alpha_{2}\
\cos\big[k\wedge
p(1-2\alpha_{1}-2\alpha_{2})\big]\nonumber\\
&&\times\frac{1}{\ln\Lambda^{2}}\left\{
\left(\ln\frac{1}{\frac{1}{\Lambda^{2}}+\frac{q\circ
q}{4}}-\ln\Delta\right)-\frac{q\circ
q}{8}\left(\frac{1}{\frac{1}{\Lambda^{2}}+\frac{q\circ
q}{4}}\right)-\Delta\ln\frac{1}{\frac{1}{\Lambda^{2}}+\frac{q\circ
q}{4}}+\Delta\ln\Delta \right\}.
\end{eqnarray}
Here, $q\equiv k+p$,
$\Delta=k^{2}\alpha_{1}(1-\alpha_{1})+p^{2}\alpha_{2}(1-\alpha_{2})+2kp\
\alpha_{1}\alpha_{2}$, and $q\circ q\equiv
-q_{\mu}\Theta^{\mu\nu}\Theta_{\nu\rho}q^{\rho}$. $\Lambda$ is the
cutoff regularization parameter. In the spirit of the  UV/IR
phenomenon \cite{minwalla}, one considers the two different
limits,  $\frac{q\circ q}{4}\gg\frac{1}{\Lambda^{2}}$ and
$\frac{q\circ q}{4}\ll\frac{1}{\Lambda^{2}}$, separately. In the
first case, $\frac{q\circ q}{4}\gg\frac{1}{\Lambda^{2}}$, one
takes the limit of the cutoff $\Lambda\to \infty$ keeping $\Theta
q$ finite. Then the anomaly vanishes due to the factor
$\frac{1}{\ln\Lambda^{2}}$ in front of the expression on the third
line of (\ref{A2-21}). On the other hand where $\frac{q\circ
q}{4}\ll\frac{1}{\Lambda^{2}}$, a finite anomaly arises due to IR
singularity; keeping $\Lambda$ large but finite, {\it i.e.}
considering the theory to be the low energy effective theory of a
fundamental theory, and taking the limit $\Theta p\to 0$, a finite
contribution from the factor
$$\frac{1}{\ln\Lambda^{2}}\ln\frac{1}{\frac{1}{\Lambda^{2}}}=1,$$
in the third line of (\ref{A2-21}) gives the nonvanishing anomaly
\begin{eqnarray}\label{A2-22}
\langle\partial_{\mu}j^{\mu}_{5}(x)\rangle=-\frac{1}{16\pi^{2}}F_{\mu\nu}\star'
\tilde{F}^{\mu\nu}+\cdots,
\end{eqnarray}
with the generalized $\star'$-product defined by
\begin{eqnarray}\label{A2-23}
f(x)\star' g(x)\equiv
f(x)\frac{\sin\left(\frac{\Theta^{\mu\nu}}{2}\overleftarrow{\partial}_{\mu}\overrightarrow{\partial}_{\nu}\right)}
{\frac{\Theta^{\mu\nu}}{2}\overleftarrow{\partial}_{\mu}\overrightarrow{\partial}_{\nu}}g(x).
\end{eqnarray}
Here, as it turns out the obtained result from the computation of
triangle, square and pentagon diagram is not gauge invariant
unless we add the contribution of infinitely many diagrams with
more and more external gauge field insertions. In \cite{neda3} the
nonplanar anomaly is calculated using the Fujikawa's path integral
method and was argued that the expression on the r.h.s. of the
above equation (\ref{A2-22}) is consistently gauge invariant only
when $F_{\mu\nu}$ and $\tilde{F}^{\mu\nu}$ are attached to an open
Wilson line as a function of the external gauge field. The
ellipses on the r.h.s. of (\ref{A2-23}) indicate the contributions
of the expansion of the Wilson line in the external gauge field.
\par
Note that the expression above remains finite in low energy
effective theory when the corresponding scale of the theory
$\Lambda$ is large comparing with $\frac{1}{|\Theta q|}$. However,
considering the gauge theory as a fundamental field theory and
letting $\Lambda$ go to infinity, the integrand of the anomaly in
the Fourier integral of (\ref{A2-21}) vanishes except for the
point $q\circ q=0$ for which the integrand becomes nonzero and
finite in agreement with the point-splitting result of Ref.
\cite{armoni}, which is then argued to be consistent with the
constraint of Eq. (\ref{A1-3}).
\par
But Fourier integral of a function which is everywhere zero and
finite at a single point (not a Dirac $\delta$-function!) is
certainly zero which is the conclusion of zero anomaly of Ref.
\cite{intriligator}. It is these conflicting conclusions that we
set out to resolve in the next section. Our position is that
because of UV/IR mixing of divergences in noncommutative theories,
the necessarily nonvanishing anomaly of the invariant axial vector
current will not emerge unless both UV and IR regularization are
introduced in the calculation. The UV regularization is introduced
by point-splitting the current; and the IR regularization is
achieved by compactification of the noncommutative directions.

\subsubsection*{3\hspace{0.3cm} IR Regularized Invariant Anomaly}
\setcounter{section}{3} \setcounter{equation}{0} In this section
we substantiate our claim that the resolution of the above paradox
is by simultaneously regularizing the currents both in IR and UV
regions. To calculate the anomaly of the divergence of the
invariant axial current $j_{5}^{\mu}$ in the noncommutative QED we
use the point-split, gauge invariant expression for the current
similar to (\ref{A2-8})
\begin{eqnarray}\label{A3-1}
j_{5}^{\mu}(x)=\lim\limits_{\epsilon\to
0}\bar{\psi}_{\alpha}(x+\frac{\epsilon}{2})\star\exp\left(-ig\int\limits_{x-\frac{\epsilon}{2}}^{x+\frac{\epsilon}{2}}
A(y)\cdot
dy\right)_{\star}\star\left(\gamma_{\mu}\gamma_{5}\right)^{\alpha\beta}\psi_{\beta}(x-\frac{\epsilon}{2}),
\end{eqnarray}
where the exponential is understood as $\star$-exponential. The
$\epsilon$-insertion is in fact a UV regulator which at the end of
calculation goes to zero.
\par
The divergence of the axial current in the limit of small
$\epsilon$ then gives
\begin{eqnarray}\label{A3-3}
\langle\partial_{\mu}j^{\mu}_{5}(x)\rangle=-ig\lim\limits_{\epsilon\to
0}\epsilon^{\nu}\langle
\bar{\psi}_{\alpha}(x+\frac{\epsilon}{2})\star F_{\mu\nu}(x)\star
\left(\gamma^{\mu}\gamma^{5}\right)^{\alpha\beta}
\psi_{\beta}(x+\frac{\epsilon}{2}) \rangle.
\end{eqnarray}
Up to the first order of perturbation theory using the action (\ref{A2-13}), the above expression reads
\begin{eqnarray}\label{A3-3a}
\lefteqn{\langle\partial_{\mu}j^{\mu}_{5}(x)\rangle=}\nonumber\\
&=&+ig^{2}\lim\limits_{\epsilon\to 0}\epsilon_{\nu}
(\gamma_{\mu}\gamma_{5})^{\alpha\beta}(\gamma_{\rho})^{\sigma\tau}\bigg\langle
\bar{\psi}_{\alpha}(x+\frac{\epsilon}{2})\stackrel{x}{\star}F^{\mu\nu}(x)\stackrel{x}{\star}
\psi_{\beta}(x-\frac{\epsilon}{2})\int d^{4}z\
\psi_{\tau}(z)\stackrel{z}{\star}\bar{\psi}_{\sigma}(z)\stackrel{z}{\star}A^{\rho}(z)\bigg\rangle\nonumber\\
&=& -ig^{2}\lim\limits_{\epsilon\to 0}\epsilon_{\nu}\int d^{4}z\
\mbox{tr}\left(S_{F}(z-x-\frac{\epsilon}{2})\gamma_{\mu}\gamma_{5}\stackrel{x}{\star}F^{\mu\nu}(x)\stackrel{x}{\star}
\stackrel{z}{\star}S_{F}(x-z-\frac{\epsilon}{2})\gamma_{\rho}\right)\stackrel{z}{\star}A^{\rho}(z),
\end{eqnarray}
where we have indicated by overwrite the relevant
$\star$-operators and further used the propagator of the massless
fermions
\begin{eqnarray}\label{A3-4}
i\left(S_{F}(x-y)\right)_{\alpha\beta}\equiv
\langle\psi_{\alpha}(x)\bar{\psi}_{\beta}(y)\rangle =
\frac{i}{(2\pi)^{2}}\
\frac{(x-y)_{\mu}\left(\gamma^{\mu}\right)_{\alpha\beta}}{\big[(x-y)^{2}-i\epsilon'\big]^{2}}.
\end{eqnarray}
Here, $\epsilon'$ arises from Feynman's $\epsilon$-prescription.
Now, it is known \cite{martin2, intriligator} that the
$\star$-operations render the integrals convergent and therefore
as a result anomaly vanishes as $\epsilon\to 0$. It is also known
that when the noncommutativity parameter $\Theta$ vanishes, the
integral diverges and the usual commutative anomaly emerges. This
indicates that under the circumstances that the $\star$-operation
is inoperative a nonzero anomaly should be obtained. We will now
make this statement precise.
\par
To do so, we introduce an IR regulator by compactifying each space
coordinates to a circle with radius $R$ that plays the role of the
IR regulator. The Fourier series expansion for the propagator
(\ref{A3-4}) therefore reads
\begin{eqnarray}\label{A3-5}
S_{F}(z)=\sum_{\vec{k}}\int\limits_{-\infty}^{+\infty}
\frac{dk_{0}}{(2\pi)^{1/2}(2R)^{3/2}} \ \frac{\KS}{k^{2}}\
e^{-ik_{0}z_{0}}e^{+i\vec{k}\cdot \vec{z}},\qquad \vec{k}\equiv
\frac{\pi\vec{n}_{k}}{R}.
\end{eqnarray}
Using this expansion in (\ref{A3-4}) we get
\begin{eqnarray}\label{A3-6}
\lefteqn{\langle\partial_{\mu}j^{\mu}_{5}(x)\rangle=-ig^{2}\lim\limits_{\epsilon\to
0}\epsilon_{\nu}\sum\limits_{\vec{p},\vec{k}}\int
\frac{dp_{0}dq_{0}}{(2\pi)(2R)^{3}}\ \mbox{tr}\left(\QS\gamma_{\mu}\gamma_{5}\PS\gamma_{\rho}\right) }\nonumber\\
&&\times\int d^{4}z \frac{e^{-iq_{0}(z-x)_{0}}e^{+i\vec{q}\cdot
(\vec{z}-\vec{x}-\vec{\epsilon}/2)}}{q^{2}}\stackrel{x}{\star}F^{\mu\nu}(x_{0},\vec{x})\stackrel{x}{\star}
\stackrel{z}{\star}\frac{e^{-ip_{0}(x-z)_{0}}e^{+i\vec{p}\cdot
(\vec{x}-\vec{z}-\vec{\epsilon}/2)}}{p^{2}}\stackrel{z}{\star}A^{\rho}(z_{0},\vec{z}).
\end{eqnarray}
The $\star$-products can be performed using
\begin{eqnarray}\label{A3-7}
e^{-i\vec{p}\cdot
\vec{z}}\stackrel{z}{\star}f(z)\stackrel{z}{\star}e^{i\vec{q}\cdot
\vec{z}}=f\left({z}-(\widehat{{p}}+\widehat{{q}})\right)\
e^{i\vec{p}\wedge\vec{q}}\ e^{-i(\vec{p}-\vec{q})\cdot \vec{z}},
\end{eqnarray}
with $\widehat{p}^{i}\equiv \frac{\Theta^{ij}p_{j}}{2}$, and
$i,j=1,2$. After an appropriate change of variable and using
$\mbox{tr}\left(\gamma_{\alpha}\gamma_{\mu}\gamma_{5}\gamma_{\beta}\gamma_{\rho}\right)
=4i\varepsilon_{\beta\rho\alpha\mu}$,
we obtain
\begin{eqnarray}\label{A3-8}
\langle\partial_{\mu}j^{\mu}_{5}(x)\rangle&=&-32g^{2}\lim\limits_{\epsilon\to
0}\epsilon_{\nu}\varepsilon_{\beta\rho\alpha\mu}\int d^{4}z
\sum\limits_{\vec{k},\vec{\ell}}\int\frac{dk_{0}d\ell_{0}}{(2\pi)(2R)^{3}}e^{-i\ell(x-z)}
\nonumber\\
&&\times \ell^{\alpha}
A^{\rho}\left({z}-\widehat{{k}}\right)F^{\mu\nu}\left({x}-\widehat{{k}}\right)
\frac{k^{\beta}e^{-i\vec{k}\cdot\vec{\epsilon}/2}}{(k+\ell)^{2}(k-\ell)^{2}}.
\end{eqnarray}
For finite noncommutative momentum $\widehat{\vec{k}}$,
$F^{\mu\nu}$ and $A^{\rho}$ will damp the
$k$-integration/sum\-mation and a finite result will arise. Taking
the zero limit of the UV regulator $\epsilon$ naively, would make
the nonplanar anomaly vanish. But, as we will see, the nonplanar
anomaly in fact receives a finite contribution from the zero modes
of the product of $A^{\rho}$ and $F^{\mu\nu}$. After performing a
series expansion of these two fields and integrating over $z$, we
get
\begin{eqnarray}\label{A3-10}
\langle\partial_{\mu}j^{\mu}_{5}(x)\rangle
&=&-32g^{2}\varepsilon_{\beta\rho\alpha\mu}
\sum\limits_{\vec{\ell},\vec{s}}\
\int\frac{d\ell_{0}ds_{0}}{(2\pi) (2R)^{3}} \ell^{\alpha}
A^{\rho}\left({\ell}\right)F^{\mu\nu}\left({s}\right)
e^{-i({s}+{\ell}) {x}}
\nonumber\\
&&\times \lim\limits_{\epsilon\to 0}
\epsilon_{\nu}\sum\limits_{\vec{k}}\int\limits_{-\infty}^{+\infty}dk_{0}\frac{k^{\beta}e^{-i{k}_{i}({\epsilon}^{i}-
\Theta^{ij} ({\ell}+{s})_{j})/2}}{(k+\ell)^{2}(k-\ell)^{2}}.
\end{eqnarray}
Let us now concentrate on the integration/summation over $k$.
Introducing the directions parallel and perpendicular to the
noncommutative  $x_1$-$x_2$ plane, $\vec{x}_{\|}=(x_{0}, x_{3})$
and $\vec{x}_{\perp}=(x_{1}, x_{2})$, we observe that the integral
is in fact UV finite for $(\vec{\ell}+\vec{s})_{\perp}\neq
\vec{0}$. In this case, the nonplanar anomaly vanishes after
taking the zero limit of UV regulator $\epsilon$. For vanishing
$(\vec{\ell}+\vec{s})_{\perp}$, however, the integral
\begin{eqnarray}\label{A3-11}
{\cal{I}}_{i}^{\ j}\equiv\lim\limits_{\epsilon\to 0}
\epsilon_{i}\sum\limits_{\vec{k}}\int\limits_{-\infty}^{+\infty}dk_{0}\frac{k^{j}e^{-i\vec{k}\cdot
\vec{\epsilon}/2}}{(k^2)^{2}}=2i\lim\limits_{\epsilon\to
0}\epsilon_{i}\frac{\partial}{\partial
\epsilon_{j}}\sum\limits_{\vec{k}}\int\limits_{-\infty}^{+\infty}
dk_{0}\ \frac{e^{-i\vec{k}\cdot\vec{\epsilon}/2}}{(k^{2})^{2}},
\end{eqnarray}
is the compactified version of the usual divergent integral
leading to anomaly. Note that comparing with the expression on the
last line of (\ref{A3-10}), we have neglected the
$\ell$-dependence in the denominator of (\ref{A3-11}). This is
because we are only interested in the UV divergent part of this
expression. In the limit of small $\epsilon$, the discrete sum can
be replaced by a logarithmical divergent integral leading to a
finite result
\begin{eqnarray}\label{A3-12}
{\cal{I}}_{i}^{\ j}\equiv iC\delta_{i}^{\ j},
\end{eqnarray}
where $C=\frac{1}{(16\pi)^{2}}$ is a numerical factor. Before
using this result a remark is in place: \par As we have mentioned
in the introduction, in \cite{armoni} the non-compactified version
of the same analysis is performed in the continuous momentum space
$q=\ell+s$. The authors have argued that the nonplanar anomaly
arises due to the UV divergence of the integral only when
$\sum_{i,j=1,2}\Theta^{ij}({\ell}+{s})_{j}=0$,\footnote{In two
dimensions $\Theta^{ij}({\ell}+{s})_{j}=0$ would lead to
$(\vec{\ell}+\vec{s})_{j}=0$, with $j=1,2$.} where in contrast to
our case $\vec{q}_{\perp}\equiv (\vec{\ell}+\vec{s})_{\perp}$ is a
{\it continuous} momentum variable. But, this is indeed a measure
zero contribution to the Lebesgues integrand over $q$ and does not
contribute to the integral. Thus the nonplanar anomaly vanishes.
In the compactified version, however, the zero mode is
well-defined and makes a finite contribution to the anomaly.
\par
Going back to our computation and inserting (\ref{A3-12}) together with a Kroeneker $\delta$-function
$\delta_{\vec{\ell}_{\perp}+\vec{s}_{\perp},\vec{0}}$ in (\ref{A3-10}), we obtain
\begin{eqnarray}\label{A3-13}
\langle\partial_{\mu}j^{\mu}_{5}(x)\rangle&=&-\frac{ig^{2}}{8\pi^{2}}\varepsilon_{j\rho\alpha\mu}
\sum\limits_{\vec{\ell},\vec{s}_{\perp}, s_{3}}\
\int\frac{d\ell_{0}ds_{0}}{(2\pi) (2R)^{3}}
\delta_{\vec{\ell}_{\perp}+\vec{s}_{\perp},\vec{0}}\ \ell^{\alpha}
A^{\rho}\left({\ell}\right)F^{\mu j}\left({s}\right)
e^{-i({s}+{\ell}){x}}.
\end{eqnarray}
The Kroeneker $\delta$-function indicates that the nonplanar anomaly exists only for vanishing discrete momentum
$(\vec{\ell}+\vec{s})_{\perp}$. Performing further the sum over $\vec{s}_{\perp}$, the total $\vec{x}_{\perp}$
dependence of the resulting expression on the r.h.s. of (\ref{A3-13}) disappears and we are left with
\begin{eqnarray}\label{A3-14}
\langle\partial_{\mu}j^{\mu}_{5}(x)\rangle
=-\frac{ig^{2}}{8\pi^{2}}\varepsilon_{j\rho\alpha\mu}
\sum\limits_{\vec{\ell},s_{3}}\ \int\frac{d\ell_{0}ds_{0}}{(2\pi)
(2R)^{3}}\ \ell^{\alpha}
A^{\rho}\left(\vec{\ell}_{\|},\vec{\ell}_{\perp}\right)F^{\mu
j}\left(\vec{s}_{\|},-\vec{\ell}_{\perp}\right)
e^{-i(\vec{s}_{\|}+\vec{\ell}_{\|})\cdot \vec{x}_{\|}}.
\end{eqnarray}
Transforming the fields back to the coordinate space, we arrive at
\begin{eqnarray}\label{A3-15}
\langle\partial_{\mu}j^{\mu}_{5}(x)\rangle
=-\frac{g^{2}}{16\pi^{2}} \frac{1}{(2R)^{2}}\int\limits_{-R}^{+R}
d^{2}y_{\perp}\
F^{\alpha\rho}(\vec{x}_{\|},\vec{y}_{\perp})\tilde{F}_{\alpha\rho}(\vec{x}_{\|},\vec{y}_{\perp}),
\end{eqnarray}
which can be interpreted as the zero modes in the Fourier expansion of the function ${\cal{A}}\equiv F\tilde{F}$
in the noncommutative coordinates $x_{\perp}$, {\it i.e.},
\begin{eqnarray}\label{A3-14b}
\langle\partial_{\mu}j^{\mu}_{5}(x)\rangle=-\frac{g^{2}}{16\pi^{2}}\ \widetilde{{\cal{A}} }(x_{\|},
p_{\perp}=0),
\end{eqnarray}
the unintegrated form of the nonplanar anomaly. The result (\ref{A3-15}) and equivalently (\ref{A3-14b}) are
gauge invariant due to cyclicity of the $\star$-product under the integral over noncommutative coordinates
$y_{\perp}$.
\par
It is clear that an integration over $\vec{x}_{\perp}$ on both
side of (\ref{A3-15}) cancels the $R$ dependence on the r.h.s.,
and the integrated form of the nonplanar anomaly become
\begin{eqnarray}\label{A3-16}
\int\limits_{-R}^{+R}
d^{2}x_{\perp}\langle\partial_{\mu}j^{\mu}_{5}(x)\rangle
&=&-\frac{g^{2}}{16\pi^{2}}\
\frac{1}{(2R)^{2}}\int\limits_{-R}^{+R}
d^{2}x_{\perp}\int\limits_{-R}^{+R} d^{2}y_{\perp}\
F^{\alpha\rho}(\vec{x}_{\|},\vec{y}_{\perp})\tilde{F}_{\alpha\rho}(\vec{x}_{\|},\vec{y}_{\perp}),
\end{eqnarray}
which survives even in $R\to \infty$ limit, {\it i.e.}
\begin{eqnarray}\label{A3-17}
\int\limits_{-\infty}^{+\infty}
d^{2}x_{\perp}\langle\partial_{\mu}j^{\mu}_{5}(x)\rangle
=-\frac{g^{2}}{16\pi^{2}}\ \int\limits_{-\infty}^{+\infty}
d^{2}y_{\perp}\
F^{\alpha\rho}(\vec{x}_{\|},\vec{y}_{\perp})\tilde{F}_{\alpha\rho}(\vec{x}_{\|},\vec{y}_{\perp}),
\end{eqnarray}
in agreement with the conclusion of \cite{armoni}.
\par
Although this result coincides with the integrated form of
nonplanar anomaly in \cite{armoni}, the unintegrated form of the
anomaly (\ref{A3-15}) in the compactified version is in fact a
novel result.\footnote{There is an alternative calculation for the
invariant anomaly \cite{ncanomalies3} using Seiberg-Witten map and
expansion of the currents in $\Theta$, which yields an
unintegrated form for the invariant anomaly. However, as
convergence properties of diagrams depend crucially on considering
all orders of the expansion, we hesitate to comment on their
result. In fact nonplanar diagram integrals would have very
different convergence properties order by order in $\Theta$ than
the expression when the phases $e^{ip\wedge q}$ are not expanded
in $\Theta$.} As we have seen the integration over noncommutative
directions $\vec{x}_{\perp}$ should be performed before taking the
limit $R\to \infty$ and this is the essential aspect of our IR
regularization. Otherwise the nonplanar anomaly of the invariant
current would vanish. And, the paradox pointed out in
\cite{armoni} would persist.
\par
In the next section, we will use the same method to calculate the
Schwinger terms in the current algebra of noncommutative currents.
\subsubsection*{4\hspace{0.3cm} Schwinger Terms}
\setcounter{section}{4} \setcounter{equation}{0} Anomalies as
quantum breakdown of classical symmetries, appearing as
contribution to the divergence of currents, are intimately related
to another manifestation of quantum violation of these symmetries,
{\it i.e.} in the additional terms to the current commutation
relations of the symmetry of the theory, the so called Schwinger
terms. \par In fact axial anomalies were first observed in the
context of the attempt to understand electroweak properties of
hadrons through the study of the corresponding current algebras
\cite{anomaly-book}.
\par
In the simplest case of massless QED with $N_{f}$ flavors, the canonical algebra of currents of {\it global
flavor symmetries}
\begin{eqnarray}\label{A4-1}
\big[j_{0}^{5(a)}(\vec{x},t),j_{0}^{b}(\vec{y},t)\big]=if^{abc}j_{0}^{5(c)}(\vec{x},t)\delta^{3}(\vec{x}-\vec{y}),
\end{eqnarray}
receives a contribution due to quantum corrections proportional to
derivatives of Dirac $\delta$-function
\begin{eqnarray}\label{A4-2}
\big[j_{0}^{5(a)}(\vec{x},t),j_{0}^{b}(\vec{y},t)\big]=if^{abc}j_{0}^{5(c)}(\vec{x},t)\delta^{3}(\vec{x}-\vec{y})+
c\delta^{ab}\varepsilon^{ijk}F_{jk}\partial_{i}\delta^{3}(\vec{x}-\vec{y}),
\end{eqnarray}
called Schwinger term. Here, the field strength tensor $F_{\mu\nu}\equiv
\partial_{\mu}A_{\nu}-\partial_{\nu}A_{\mu}$ and $f^{abc}$ are the structure functions of the relevant groups, and the
coefficient $c$ is, what in the affine algebra context is called central charge. It is well-known that anomalies
and Schwinger terms are related through the so called descent equations. However, a more direct connection
between the equal-time commutation (ETC) relation and the anomaly is illustrated most easily in a detailed
calculation of the anomaly in perturbation theory [see Appendix A].
\par
In the case of noncommutative QED, as we have two distinct
currents, the invariant currents $j^{\mu}$ or $j^{\mu}_{5}$
(\ref{A2-17}) and the covariant currents $J^{\mu}$ and
$J^{\mu}_{5}$ (\ref{A2-18}), would have a number of different
Schwinger terms. The Schwinger term appearing in the commutation
relation of two covariant currents involves planar diagrams,
whereas that corresponding commutator of a covariant and an
invariant current involves nonplanar diagrams.
\par
As an example\footnote{The full algebra of noncommutative currents
will be presented elsewhere \cite{neda5}.}, we calculate the
equal-time commutation (ETC) relation between $J^{0,5}$ and
$J^{0}$. It consists of a canonical and a Schwinger term. The
canonical part is easily seen to be
\begin{eqnarray}\label{A4-3aa}
\big[J^{0}_{5}(\vec{x},t),J^{0}(\vec{y},t))\big]\Bigg|_{\mbox{\small{can.}}}&=&+2i\int
\frac{d^{4}p}{(2\pi)^{4}}\ \frac{d^{4}q}{(2\pi)^{4}}
\frac{d^{4}\ell}{(2\pi)^{4}}\
\psi_{\beta}({\ell}+{p}+{q})\bar{\psi}_{\alpha}({\ell})
\left(\gamma_{0}\gamma_{5}\right)^{\alpha\beta}
\nonumber\\
&&\times e^{-i(p+q)_{0}t}\
e^{+i\vec{p}\cdot\vec{x}}e^{+i\vec{q}\cdot\vec{y}}\
e^{-i\vec{\ell}\wedge\left(\vec{p}+\vec{q}\right)}\
\sin\left(\vec{p}\wedge \vec{q}\right),
\end{eqnarray}
which after some algebras becomes
\begin{eqnarray}\label{A4-3a}
\big[J^{0}_{5}(\vec{x},t),J^{0}(\vec{y},t))\big]\Bigg|_{\mbox{\small{can.}}}=
\bigg[J_{5}^{0}(\vec{x},t)-J_{5}^{0}(\vec{y},t)\bigg]\star
\delta^{3}\left(\vec{x}-\vec{y}\right),
\end{eqnarray}
and in momentum space
\begin{eqnarray}\label{A4-3b}
\big[J^{0}_{5}(\vec{p},t),J^{0}(\vec{q},t)\big]\Bigg|_{\mbox{\small{can.}}}=
+2i\sin\left(\vec{p}\wedge\vec{q}\right)J_{5}^{0}\left(\vec{p}+\vec{q},t\right).
\end{eqnarray}
As for the corresponding Schwinger term, we have to calculate the vev of the ETC relation, which is related to
the planar anomaly arising from the covariant current \cite{neda1}.\footnote{In Appendix A we have shown that in
noncommutative QED, since the current appearing in the Lagrangian density is the covariant vector current
$J_{\mu}$, the calculation of the anomaly corresponding to the covariant axial vector current $J_{\mu}^{5}$
involves the ETC of a covariant axial vector current with a covariant vector current, $[J^{0}_{5}(\vec{x},t),
J^{0}(\vec{y},t)]$, whereas the calculation of the nonplanar anomaly corresponding to the invariant axial vector
current $j_{\mu}^{5}$ involves the ETC of an invariant axial vector current with a covariant current,
$[j^{0}_{5}(\vec{x},t), J^{0}(\vec{y},t)]$.} Using a point-splitting regularization and after some algebra we
get
\begin{eqnarray}\label{A4-3ell}
\lefteqn{\langle\big[J^{0}_{5}(\vec{x},t),J^{0}(\vec{y},t))\big]\rangle=}\nonumber\\
&=&+32ig\lim\limits_{\epsilon\to 0}\ \varepsilon_{\beta\rho\alpha 0}\ \int d^{4}z\int
\frac{d^{4}\ell}{(2\pi)^{4}}\ell^{\alpha}A^{\rho}(z)e^{+i\ell_{0}\left(t-z_{0}\right)}\
e^{-i\vec{\ell}\cdot(\vec{x}-\vec{z})}\int\frac{d^{4}k}{(2\pi)^{4}}\
\frac{k^{\beta}\ e^{+i\vec{k}\cdot{\vec{\epsilon}}/{2}}}{(k+\ell)^{2}(k-\ell)^{2}}\nonumber\\
&&\times\left(\delta^{3}(\vec{x}-\vec{y}+\frac{\vec{\epsilon}}{2}+\widehat{\vec{\ell}})-
\delta^{3}(\vec{y}-\vec{x}+\frac{\vec{\epsilon}}{2}+\widehat{\vec{\ell}})\right),
\end{eqnarray}
at one-loop order, with $\widehat{{\ell}}$ defined by
$\widehat{{\ell}}^{i}\equiv \frac{\Theta^{ij}\ell_{j}}{2}$. As we
are interested in the UV behavior of the $k$-integration, we can
neglect the $\ell$ dependence in the denominator, and arrive after
some manipulations at
\begin{eqnarray}\label{A4-3c}
\langle\big[J^{0}_{5}(\vec{x},t),J^{0}(\vec{y},t))\big]\rangle
&=&-64ig\lim\limits_{\epsilon\to 0}\ \varepsilon_{\beta\rho\alpha
0}\ \int d^{4}z\
\partial^{\alpha}A^{\rho}(z)e^{+i\ell_{0}\left(t-z_{0}\right)}\
e^{-i\vec{\ell}\cdot(\vec{x}-\vec{z})}\
\int\frac{d^{4}k}{(2\pi)^{4}}\frac{k^{\beta}\
e^{+i\vec{k}\cdot{\vec{\epsilon}}/{2}}}{(k^{2})^{2}}\nonumber\\
&&\times \int\frac{d^{3}q}{(2\pi)^{3}}\
e^{+i\vec{q}\cdot(\vec{x}-\vec{y})}\left(\sin\left(\vec{q}\wedge\vec{\ell}\right)+
\frac{\vec{\epsilon}\cdot\vec{q}}{2}\cos\left(\vec{q}\wedge\vec{\ell}\right)\right).
\end{eqnarray}
The integral over $k$ diverges linearly in the limit $\epsilon\to
0$. Taking an appropriate symmetric limit, the term proportional
to $\sin\left(\vec{q}\wedge\vec{\ell}\right)$ cancels out and we
are left with the term proportional to
$\cos\left(\vec{q}\wedge\vec{\ell}\right)$. Using the result from
previous section in performing the $k$-integration, we get
\begin{eqnarray*}\label{A4-3e}
\lefteqn{\langle\big[J^{0}_{5}(\vec{x},t),J^{0}(\vec{y},t))\big]\rangle
=}\nonumber\\
&&=-\frac{g}{8\pi^{2}}\varepsilon^{ijk}\int d^{4}z\
\frac{d^{4}\ell}{(2\pi)^{4}}\ \frac{d^{3}q}{(2\pi)^{3}}\
\partial_{k}A_{j}(z) q_{i}\ e^{+i\ell_{0}\left(t-z_{0}\right)}\
e^{-i\vec{\ell}\cdot(\vec{x}-\vec{z})}\ e^{i\vec{q}\cdot
(\vec{x}-\vec{y})}\ \cos(\vec{q}\wedge\vec{\ell}).
\end{eqnarray*}
The Schwinger term corresponding to the current algebra of the
covariant currents then becomes
\begin{eqnarray}\label{A4-3f}
\langle\big[J^{0}_{5}(\vec{x},t),J^{0}(\vec{y},t))\big]\rangle
=+\frac{ig}{16\pi^{2}}\varepsilon^{ijk}\bigg[\partial_{k}A_{j}(\vec{x},t)\star\partial_{i}\delta^{3}(\vec{x}-\vec{y})
+\partial_{i}\delta^{3}(\vec{x}-\vec{y})\star
\partial_{k}A_{j}(\vec{x},t)\bigg].
\end{eqnarray}
In momentum space, it reads
\begin{eqnarray}\label{A4-3g}
\langle\big[J^{0}_{5}(\vec{p},t),J^{0}(\vec{q},t)\big]\rangle
=-\frac{ig}{8\pi^{2}}\cos\left(\vec{p}\wedge
\vec{q}\right)\varepsilon^{ijk}p_{k}q_{i}A_{j}\left(\vec{p}+\vec{q},t\right).
\end{eqnarray}
Combining the Schwinger term (\ref{A4-3f}) [(\ref{A4-3g})] with
the canonical term (\ref{A4-3a}) [(\ref{A4-3b})], the ETC of
covariant currents in the coordinate space is given by
\begin{eqnarray}\label{A4-4a}
\big[J^{0}_{5}(\vec{x},t),J^{0}(\vec{y},t))\big]&=&\left(J^{0}_{5}(\vec{x},t)-J^{0}_{5}(\vec{y},t)
\right)\star\delta^{3}(\vec{x}-\vec{y})\nonumber\\
&&+\frac{ig}{32\pi^{2}}\varepsilon^{ijk}\bigg[F_{kj}(\vec{x},t)\star\partial_{i}\delta^{3}(\vec{x}-\vec{y})
+\partial_{i}\delta^{3}(\vec{x}-\vec{y})\star
F_{kj}(\vec{x},t)\bigg],
\end{eqnarray}
and in momentum space by
\begin{eqnarray}\label{A4-3h}
\lefteqn{\big[J^{0}_{5}(\vec{p},t),J^{0}(\vec{q},t)\big]
=}\nonumber\\
&&=+2i\sin\left(\vec{p}\wedge \vec{q}\right)J_{5}^{0}(
\vec{p}+\vec{q},t)-\frac{ig}{8\pi^{2}}\cos\left(\vec{p}\wedge
\vec{q}\right)\varepsilon^{ijk}p_{k}q_{i}A_{j}\left(\vec{p}+\vec{q},t\right).
\end{eqnarray}
The ETC of covariant currents up to second order in
$\Theta$-expansion are calculated  recently in \cite{banerjee-ETC}
using an appropriate Seiberg-Witten map.
\par
The Schwinger term of the commutator of an invariant current and a
covariant current is related to the invariant current divergence
and its anomaly, and as we have mentioned before involves
nonplanar diagrams.
\par
The ETC relation between  the zero component of an invariant axial
vector current $j^{0}_{5}$ and covariant vector current $J^{0}$
$$ \big[j^{0}_{5}(\vec{x},t),
J^{0}(\vec{y},t))\big],
$$
involves only an anomalous term. This Schwinger term is calculated in a similar manner to the anomaly of section
3 using a point split regularization in three spatial directions. In the first order of perturbative expansion
it reads
\begin{eqnarray*}\label{A4-4}
\lefteqn{\hspace{-0.3cm}\langle\big[j^{0}_{5}(\vec{x},t),J^{0}(\vec{y},t))\big]\rangle=}\nonumber\\
=&&\hspace{-0.5cm}+g\lim\limits_{\epsilon\to 0}\bigg[
\delta^{3}(\vec{y}-\vec{x}+\frac{\vec{\epsilon}}{2})\stackrel{x}{\star}\stackrel{y}{\star}
\psi_{\beta}(\vec{x}+\frac{\vec{\epsilon}}{2},t)\
\bar{\psi}_{\alpha}(y)\left(\gamma_{0}\gamma_{5}\right)^{\alpha\beta}
\int d^{4}z\
\psi_{\tau}(z)\stackrel{z}{\star}\bar{\psi}_{\sigma}(z)\stackrel{z}{\star}(\gamma_{\rho})^{\sigma\tau}A^{\rho}(z)
\nonumber\\
&&-\psi_{\beta}(y)\
\bar{\psi}_{\alpha}(\vec{x}-\frac{\vec{\epsilon}}{2},t)\left(\gamma_{0}\gamma_{5}\right)^{\alpha\beta}
\stackrel{x}{\star}\stackrel{y}{\star}\delta^{3}(\vec{x}-\vec{y}+\frac{\vec{\epsilon}}{2})
\int d^{4}z\
\psi_{\tau}(z)\stackrel{z}{\star}\bar{\psi}_{\sigma}(z)\stackrel{z}{\star}(\gamma_{\rho})^{\sigma\tau}A^{\rho}(z)
\bigg]\bigg|_{y_{0}=x_{0}}\nonumber\\
=&&\hspace{-0.5cm}+g\lim\limits_{\epsilon\to 0}\bigg[\int d^{4}z\
\delta^{3}(\vec{y}-\vec{x}+\frac{\vec{\epsilon}}{2})\stackrel{x}{\star}\stackrel{y}{\star}
\
\mbox{tr}\left(S_{F}(z-y)\gamma_{0}\gamma_{5}\stackrel{z}{\star}S_{F}(\vec{x}-\vec{z}+
\frac{\vec{\epsilon}}{2},t-z_{0})\gamma_{\rho}
\right)\stackrel{z}{\star}A^{\rho}(z)\nonumber\\
&&+\int d^{4}z\
\mbox{tr}\left(S_{F}(\vec{z}-\vec{x}+\frac{\vec{\epsilon}}{2},z_{0}-t)\gamma_{0}\gamma_{5}\stackrel{z}{\star}
S_{F}(y-z)\gamma_{\rho}\right)\stackrel{x}{\star}\stackrel{y}{\star}\delta^{3}(\vec{x}-\vec{y}
+\frac{\vec{\epsilon}}{2})\stackrel{z}{\star}A^{\rho}(z)\bigg]\bigg|_{y_{0}=x_{0}},
\end{eqnarray*}
with the massless fermion propagator given in (\ref{A3-4}). As in
the case of nonplanar anomaly, under certain circumstances the
$\star$-product becomes inoperative and a nonvanishing Schwinger
term emerges. To show this we perform an IR regularization by
compactifying each space coordinate to a circle with radius $R$.
After expanding the above result in the Fourier series and some
straightforward manipulations we get
\begin{eqnarray}\label{A4-5}
\lefteqn{\langle\big[j^{0}_{5}(\vec{x},t),J^{0}(\vec{y},t))\big]\rangle=}\nonumber\\
&&\hspace{-0.7cm}=+32ig\lim\limits_{\epsilon\to 0}\varepsilon_{\beta\rho\alpha 0}\ \int d^{4}z\
\sum\limits_{\vec{k},\vec{\ell}}\int \frac{dk_{0}d\ell_{0}}{(2\pi)(2R)^{3}} e^{+i\ell_{0}\left(t-z_{0}\right)}\
e^{-i\vec{\ell}\cdot(\vec{x}-\vec{z})}\ell^{\alpha}A^{\rho}\left({z}-
\widehat{{k}}\right)\frac{k^{\beta}e^{+i\vec{k}\cdot\vec{\epsilon}/2}}{(k+\ell)^{2}(k-\ell)^{2}}\nonumber\\
&&\times
\left(\delta^{3}(\vec{y}-\vec{x}+\frac{\vec{\epsilon}}{2}+\widehat{\vec{k}})-
\delta^{3}(\vec{x}-\vec{y}+\frac{\vec{\epsilon}}{2}
-\widehat{\vec{k}})\right).
\end{eqnarray}
As expected this result is very similar to the result obtained for
the nonplanar anomaly in the previous section [see Eq.
(\ref{A3-8})]. Here, as in the previous case, for finite
 $\widehat{\vec{k}}$ the integration/summation over $k$
 remains finite and the Schwinger term vanishes by taking naively the limit $\epsilon\to
 0$. To show this, let us give the Schwinger term in the Fourier space,
 \begin{eqnarray}\label{A4-6}
\langle\big[j^{0}_{5}(\vec{p},t),J^{0}(\vec{q},t)\big]\rangle
&=&-64 g\ \epsilon_{\beta\rho\alpha
0}\sum\limits_{\vec{\ell}}\int\frac{d\ell_{0}}{(2\pi)^{1/2}(2R)^{3/2}}
e^{i\ell_{0}t}\delta_{\vec{p}+\vec{q}+\vec{\ell},\vec{0}}\
\ell^{\alpha}A^{\rho}(-\ell)\nonumber\\
&&\times \lim\limits_{\epsilon\to 0}\ \sin\left(\frac{\vec{q}\cdot
\vec{\epsilon}}{2}\right)
\sum\limits_{\vec{k}}\int\limits_{-\infty}^{+\infty}dk_{0}\frac{k^{\beta}e^{+i{k}_{i}({\epsilon}^{i}-
\Theta^{ij}{(q+\ell)}_{j})/2}}{(k+\ell)^{2}(k-\ell)^{2}}.
 \end{eqnarray}
As in the case of anomaly, the integral/sum over $k$ is finite due
to the damping effect of the gauge field $A^{\rho}$, as long as
$({\vec{\ell}+\vec{q})_{\perp}}$ is nonzero. In this case the
Schwinger term vanishes in the limit $\epsilon\to 0$. For
$({\vec{\ell}+\vec{q})_{\perp}}=\vec{0}$, however, a finite
contribution will arise. Its finite value will be calculated in
the following. We expand the $\sin$ for small $\epsilon$, and
manipulate the $k$ integration/summation as in (\ref{A3-11}).
Using the result from (\ref{A3-12}), (\ref{A4-6}) reads
 \begin{eqnarray}\label{A4-7}
\langle\big[j^{0}_{5}(\vec{p},t),J^{0}(\vec{q},t)\big]\rangle
&=&+\frac{i g}{8\pi^{2}}\
\epsilon_{ijk}\sum\limits_{\vec{\ell}}\int\frac{d\ell_{0}}{(2\pi)^{1/2}(2R)^{3/2}}
e^{i\ell_{0}t}\delta_{\vec{p}+\vec{q}+\vec{\ell},\vec{0}}
\delta_{\vec{\ell}_{\perp}+\vec{q}_{\perp},\vec{0}}\
q^{i}\ell^{k}A^{j}(-\ell),\nonumber\\
&=&-\frac{ig}{8\pi^{2}}\frac{1}{(2R)^{3/2}}\
\epsilon_{ij3}q^{i}p^{3}A^{j}(\vec{p}+\vec{q},t)\delta_{\vec{p}_{\perp,\vec{0}}},
\end{eqnarray}
where the Kroeneker $\delta$-function $\delta_{\vec{\ell}_{\perp}+\vec{q}_{\perp},\vec{0}}$ is inserted to
indicate that the Schwinger term exists only when $({\vec{\ell}+\vec{q})_{\perp}}=\vec{0}$. Now going back to
the coordinate space, and after some algebraic manipulations similar to the anomaly case, the first quantum
correction to ETC of two different noncommutative currents in the coordinate space is given by
\begin{eqnarray}\label{A4-8}
\big[j^{0}_{5}(\vec{x},t),J^{0}(\vec{y},t)\big]=+
\frac{1}{(2R)^{2}}\ \frac{ig}{8\pi^{2}}\
\partial_{3}^{x}\delta(x_{3}-y_{3})\varepsilon_{ij3}\partial^{i}A^{j}(\vec{y}_{\perp},
x_{3},t),
\end{eqnarray}
and in momentum space,
 \begin{eqnarray}\label{A4-9}
\big[j^{0}_{5}(\vec{p},t),J^{0}(\vec{q},t)\big]=-\frac{ig}{8\pi^{2}}\
\frac{1}{(2R)^{3/2}}\
\epsilon_{ij3}q^{i}p^{3}A^{j}(\vec{p}+\vec{q},t)\delta_{\vec{p}_{\perp,\vec{0}}}.
\end{eqnarray}
Note that the $R$-dependence on the r.h.s. of (\ref{A4-8}) is also shared by the result (\ref{A3-15}) of the
unintegrated form of the nonplanar (invariant) anomaly. In the discussion following (\ref{A3-15}) and leading to
the integrated form of the anomaly from (\ref{A3-17}), we have shown that the $R$-dependence disappears upon
integration over two noncommutative directions. In the case of current algebras, it is also possible to compare
the results (\ref{A4-3h}) for the ETC of two covariant currents and (\ref{A4-8}) for the ETC of an invariant and
a covariant current to check the consistency of the result (\ref{A4-8}). To do this, it is enough to integrate
these two relations over three spatial coordinates $x$ and $y$, involving also the noncommutative directions and
to show after taking the limit $R\to\infty$ that\footnote{Since the r.h.s. of (\ref{A4-8}) does not depend on
noncommutative coordinates $\vec{x}_{\perp}=(x_{1}, x_{2})$, its $R$-dependence is removed upon integration over
these noncommutative coordinates. }
$$ [Q_{5}^{inv.}, Q^{cov.}]=[Q_{5}^{cov.},Q^{cov.}],$$
\begin{eqnarray*}
\hspace{-2.5cm}\mbox{with}\qquad\qquad\qquad\qquad Q_{5}^{inv.}\equiv \int d^{3}x\
j^{0}_{5}(\vec{x},t),\qquad\mbox{}\qquad Q_{5}^{cov.}\equiv \int d^{3}x\ J^{0}_{5}(\vec{x},t),
\end{eqnarray*}
and $$Q^{cov.}\equiv \int d^{3}x\ J^{0}(\vec{x},t).$$
\par
As the case of commutator of two invariant currents is more involved and its physical significance less clear to
us, we will postpone it to a more detailed publication \cite{neda5}.

\subsubsection*{5\hspace{0.3cm} Discussion}
\setcounter{section}{4} \setcounter{equation}{0} The lesson of
UV/IR mixing in noncommutative field theories is that UV
regularization of the theory does not render the theories
consistent at the limit of zero momentum and the UV singularities
reappear as IR singularities. In this work we have reexamined the
invariant (nonplanar) anomaly and Schwinger terms in
noncommutative $U(1)$ gauge theories with care. It became clear
that the resolution of the question of nonzero integrated anomaly
requires an IR cutoff, such as the compactification length. The
anomaly and also the Schwinger term vanish as the compactification
length is removed. This agrees with the observation of
Intriligator and Kumar \cite{intriligator} that based on general
properties of nonplanar diagram, argued that for finite
noncommutativity parameter there is no UV divergence and hence no
anomaly. These arguments were supported in analogy with string
theory where the nonplanar anomalies vanish by Green-Schwarz (GS)
mechanism of anomaly cancellation.
\par
If we integrate the expression before removing the IR cutoff we
obtain the integrated anomaly in agreement with the covariant
anomaly. The result is very similar to the standard anomaly where
it naively becomes zero if the UV divergence is not carefully
handled. We also observe that the observation of Armoni, Lopez and
Theisen \cite{armoni} about nonzero result when $|\Theta p|=0$
acquires meaning with finite IR cutoff where the Fourier integral
becomes Fourier series and there is no zero measure difficulty. It
is as if a finite charge is evenly distributed over the space
giving zero density but still being totally nonzero. Therefore
there is no ambiguity or inconsistency if the IR regulator is
taken to its limit {\it after} the integration over noncommutative
space coordinates are taken.
\par
What is interesting is that the expressions for nonplanar
anomalous term turn out to be independent of the noncommutative
coordinates. Although we start with a local object, the divergence
of a current, the result is an integral over the noncommutative
part of the space washing all remnants of locality.

We can look at this phenomenon from two different points; pure field theory and String theory. Looking from the
field theory side the formula (3.8) shows that although we are considering divergence of the invariant current
at a point, $x$, the expression involves points which are away  by amount $\Theta k$. Upon integration over $k$
this shift in position will cover the whole space without damping and makes the result independent of the
position $x$ where we are calculating the divergence. This is how large momentum integration which is a UV
effect is reflected in the infinite wavelength, a constant term IR effect. Physically, large momentum states of
the particle circulating in the loop are extended in the direction perpendicular to the momentum. This extension
is the reason that the range of the nonlocality inherited from non commutativity extends to infinity, {\it i.e.}
UV/IR mixing.

When the space is finite, this extension covers the compact
dimension globally. This brings us to the stringy point of view.
Such large extension reveals the stringy nature of noncommutative
theories. The particle circulating in the loop is extended to the
extent that it wraps around the compact direction and forms a
winding state. The contribution of such winding states survives
and leads to finite anomaly. Authors of Ref. \cite{intriligator}
have suggested that anomaly cancels due to hidden GS-mechanism,
{\it i.e.} via formation of closed string modes. This picture is
complete only when the noncommutative plane is not compact. In the
compact case, the extra winding states exist and couple to the
states of the gauge theory. They develop IR poles proportional to
$R^2$. As expected this contribution vanishes as the
compactification length becomes infinitely large. At this limit
the winding states decouple and their contribution to all
processes including anomaly become zero.

Let us examine why this result can be interpreted as exchange of
winding states. Winding states have masses proportional to $R^2$.
Hence their contribution includes a propagator factor $1/((\Theta
p)^2-R^2)$ which becomes $1/R^2$ since the total noncommutative
momentum is zero. This factor is present in the final result. The
anomaly is constant in the noncommutative plane, which we
attribute to the global nature of the winding states that are by
nature nonlocal.

These observations explain why we need to resort to IR cutoff to
extract the anomaly like the axial anomaly which needs careful UV
regularization to be revealed. Similar points concerning
contribution of winding states to nonplanar Feynman diagrams are
discussed in the context of noncommutative field theories at
nonzero temperature \cite{temperature}.

Winding states have similar manifestations in the Schwinger terms.
In (\ref{A4-8}) we see that the anomalous contribution to the
commutators of the invariant axial current and covariant current
becomes independent of the noncommutative coordinates of invariant
axial current similar to the absence of noncommutative coordinates
in invariant anomaly. Hence we conclude that winding string states
are responsible for the Schwinger terms that also vanish at the
infinite $R$ limit supported by general theorems connecting the
axial anomalies and Schwinger terms. We observe their validity
also in the noncommutative case as is shown by direct
calculations.

It is interesting to find how much of string theory is hidden in
noncommutative theories and to what extent the spectrum and
interactions of the underlying string theory can be extracted from
them.

There are physical issues to be resolved concerning the decay of
the Goldstone boson of the broken (global) symmetry. At first
sight on can observe that the decay must be enhanced in the
commutative directions. The details of the enhancement and its
observable consequences are relegated to future publication. Since
quantum effects are responsible for strong breaking of Lorentz
invariance the meaning of the particle states have to be
reexamined in the noncommutative cases which is needed for the
interpretation of the results  in particular the meaning of  poles
in the zero momentum in the noncommutative plane. We refer
discussion on such consequences and other physical consequences to
future publications.

\subsubsection*{6\hspace{0.3cm} Acknowledgments}
This work is supported in part by Iranian TWAS chapter based at
ISMO.
\subsubsection*{Appendix A}
\begin{appendix}
\setcounter{section}{1} \setcounter{equation}{0} In this Appendix we will show the connection between the
equal-time commutation (ETC) relation and the anomaly of the axial vector currents using perturbative expansion.
In the ordinary commutative $U(1)$ gauge theory the action is given by
\begin{eqnarray}
S\equiv \int d^{4}x\ \bigg[\bar{\psi}(x)(i\PARTIALS-m)\psi(x)-g\ j_{\mu}(x)A^{\mu}(x)\bigg],
\end{eqnarray}
where $j_{\mu}\equiv \bar{\psi}\gamma_{\mu}\psi$. Using this action, the v.e.v. of the divergence of the axial
vector current $j_{\mu}^{5}\equiv \bar{\psi}\gamma_{\mu}\gamma_{5}\psi$ is given in a perturbative expansion by
\begin{eqnarray}
\langle\partial_{\mu}j^{\mu}_{5}(x)\rangle= -\frac{g^{2}}{2}\int d^{4}y\ d^{4}z\
\partial_{\mu}\Gamma^{\mu\nu\rho}(x,y,z)\ A_{\nu}(y)A_{\rho}(z),
\end{eqnarray}
with the vertex function
\begin{eqnarray}
\Gamma_{\mu\nu\rho}(x,y,z)\equiv\langle T\left(j_{\mu}^{5}(x)j_{\nu}(y)j_{\rho}(z)\right)\rangle.
\end{eqnarray}
Building the divergence of $\Gamma_{\mu\nu\rho}(x,y,z)$ with respect to $x$, and using the definition of the
$T$-ordered product, we arrive at
\begin{eqnarray}
\partial^{\mu}_{x}\Gamma_{\mu\nu\rho}(x,y,z)&=&\partial^{\mu}_{x}\langle
T\left(j_{\mu}^{5}(x)j_{\nu}(y)j_{\rho}(z)\right)\rangle\nonumber\\
&=&\langle T\left((\partial^{\mu}j_{\mu}^{5}(x))\ j_{\nu}(y)j_{\rho}(z)\right)\rangle+\delta(x^{0}-y^{0})\langle
T\left(\big[j_{0}^{5}(x), j_{\nu}(y)\big]\ j_{\rho}(z)\right)\rangle\nonumber\\
&&+\delta(x^{0}-z^{0})\langle T\left(j_{\nu}(y)\ \big[j_{0}^{5}(x), j_{\rho}(z)\big]\right)\rangle.
\end{eqnarray}
In the last two terms, there appears two ETC relations $\big[j_{0}^{5}(\vec{x},t), j_{\nu}(\vec{y},t)\big]$ and
$\big[j_{0}^{5}(\vec{x},t), j_{\rho}(\vec{z},t)\big]$. This is a formal connection between the anomaly and the
current algebra in the commutative $U(1)$-gauge theory.
\\
In the following, we will use the same argument to show the connection between the anomaly corresponding to the
covariant axial vector current $J_{\mu}^{5}$ with the ETC of a covariant axial vector current and a covariant
vector current, $[J_{0}^{5}(\vec{x},t), J_{0}(\vec{y},t)]$, and the nonplanar anomaly corresponding to the
invariant axial vector current $j_{\mu}^{5}$ with the ETC of an invariant axial vector current and a covariant
current, $[J_{0}^{5}(\vec{x},t), J_{0}(\vec{y},t)]$.
\par
The action of the noncommutative $U(1)$ gauge theory with fermions in the fundamental representation is given by
\begin{eqnarray}
S_{\mbox{\small{non-com}}}\equiv \int d^{4}x\ \bigg[\bar{\psi}(x)(i\PARTIALS-m)\psi(x)-g\ J_{\mu}(x)\star
A^{\mu}(x)\bigg],
\end{eqnarray}
where $J_{\mu}\equiv \psi_{\beta}\star\bar{\psi}_{\alpha}(\gamma_{\mu})^{\alpha\beta}$ is the covariant current.
According to the above perturbative argument the divergence of the covariant axial vector current
$J_{\mu}^{5}\equiv \psi_{\beta}\star\bar{\psi}_{\alpha}(\gamma_{\mu}\gamma_{5})^{\alpha\beta}$ is given by
\begin{eqnarray}
\langle\partial_{\mu}J^{\mu}_{5}(x)\rangle=-\frac{g^{2}}{2}\int d^{4}y\ d^{4}z\
\partial_{\mu}\Gamma^{\mu\nu\rho}_{cov.}(x,y,z)\ A_{\nu}(y)A_{\rho}(z),
\end{eqnarray}
where the vertex function of one {\it covariant} axial vector currents and two {\it covariant} vector currents
\begin{eqnarray}
\Gamma_{\mu\nu\rho}^{cov.}(x,y,z)\equiv\langle T\left(J_{\mu}^{5}(x)J_{\nu}(y)J_{\rho}(z)\right)\rangle.
\end{eqnarray}
Using again the definition of the $T$-ordered product, the partial derivation of $\Gamma_{\mu\nu\rho}^{cov.}$
with respect to $x$ reads
\begin{eqnarray}\label{AA8}
\partial^{\mu}_{x}\Gamma_{\mu\nu\rho}^{cov.}(x,y,z)&=&\partial^{\mu}_{x}\langle
T\left(J_{\mu}^{5}(x)J_{\nu}(y)J_{\rho}(z)\right)\rangle\nonumber\\
&=&\langle T\left((\partial^{\mu}J_{\mu}^{5}(x))\ J_{\nu}(y)J_{\rho}(z)\right)\rangle+\delta(x^{0}-y^{0})\langle
T\left(\big[J_{0}^{5}(x), J_{\nu}(y)\big]\ J_{\rho}(z)\right)\rangle\nonumber\\
&&+\delta(x^{0}-z^{0})\langle
T\left(J_{\nu}(y)\ \big[J_{0}^{5}(x), J_{\rho}(z)\big]\right)\rangle.
\end{eqnarray}
On the r.h.s., there appears the ETC relations of a {\it covariant} axial vector current and a {\it covariant}
vector current $\big[J_{0}^{5}(\vec{x},t), J_{\nu}(\vec{y},t)\big]$ and $\big[J_{0}^{5}(\vec{x},t),
J_{\rho}(\vec{z},t)\big]$.
\par
In contrast, by building the divergence of the invariant axial vector current $j_{\mu}^{5}\equiv \bar{\psi}\
\gamma_{\mu}\gamma_{5}\star\psi$, we arrive at
\begin{eqnarray}
\langle\partial_{\mu}j^{\mu}_{5}(x)\rangle=-\frac{g^2}{2}\int d^{4}y\ d^{4}z\
\partial_{\mu}\Gamma^{\mu\nu\rho}_{inv.}(x,y,z)\ A_{\nu}(y)A_{\rho}(z),
\end{eqnarray}
with the vertex function of one {\it invariant} axial vector current and two {\it covariant} vector
currents
\begin{eqnarray}
\Gamma_{\mu\nu\rho}^{inv.}(x,y,z)\equiv\langle T\left(j_{\mu}^{5}(x)J_{\nu}(y)J_{\rho}(z)\right)\rangle.
\end{eqnarray}
The divergence of this vertex function involves the ETC of an {\it invariant} axial vector current and a {\it
covariant} vector current $\big[j_{0}^{5}(\vec{x},t), J_{\nu}(\vec{y},t)\big]$ and $\big[j_{0}^{5}(\vec{x},t),
J_{\rho}(\vec{z},t)\big]$
\begin{eqnarray}\label{AA11}
\partial^{\mu}_{x}\Gamma_{\mu\nu\rho}^{inv}(x,y,z)&=&\partial^{\mu}_{x}\langle
T\left(j_{\mu}^{5}(x)J_{\nu}(y)J_{\rho}(z)\right)\rangle\nonumber\\
&=&\langle T\left((\partial^{\mu}j_{\mu}^{5}(x))\ J_{\nu}(y)J_{\rho}(z)\right)\rangle+\delta(x^{0}-y^{0})\langle
T\left(\big[j_{0}^{5}(x), J_{\nu}(y)\big]\ J_{\rho}(z)\right)\rangle\nonumber\\
&&+\delta(x^{0}-z^{0})\langle T\left(J_{\nu}(y)\ \big[j_{0}^{5}(x), J_{\rho}(z)\big]\right)\rangle.
\end{eqnarray}
To find a more direct relation between the anomaly and the Schwinger terms, one has to calculate the full
commutator algebra of noncommutative currents and insert it into the r.h.s. of the above equations (\ref{AA8})
and (\ref{AA11}). We will postpone this calculation to a more detailed publication \cite{neda5}.

\end{appendix}

\end{document}